\documentstyle[aps,preprint,epsf]{revtex}

\def\blanklinethree{}
\def\blanklinefour{}
\def\ltap{\ \raise.3ex\hbox{$<$\kern-.75em\lower1ex\hbox{$\sim$}}\ }
\def\gtap{\ \raise.3ex\hbox{$>$\kern-.75em\lower1ex\hbox{$\sim$}}\ }

\def\Lc{\Lambda_{conf}}

\def\smt{$SU(3)\times SU(2)\times U(1)$}
\def\ie{{\it i.e.}}

\def\eg{{\it e.g.}}

\def\CN{{\cal N}}
\def\CO{{\cal O}}

\def\TeV{\,{\rm TeV}}
\newcommand{\be}[1]{\begin{equation}\label{#1}}
\newcommand\ee{\end{equation}}
\newcommand{\eref}[1]{(\ref{#1})}
\newcommand{\Eref}[1]{Eq.~(\ref{#1})}
\newcommand{\SEC}[1]{Sec.~\ref{sec:#1}}
\newcommand{\SSEC}[1]{Sec.~\ref{ssec:#1}}

\def\half{{1\over2}}
\def\rarr{\rightarrow}
\def\none{$\CN=1$}
\def\ie{{\it i.e.}}

\def\eg{{\it e.g.}}
\def\G{\tilde{G}}
\def\F{\sqrt{F}}
\def\susy{supersymmetry}
\newcommand{\vev}[1]{\langle#1\rangle}
\def\lsim{\alt}

\def\susic{supersymmetric}
\def\ssb{supersymmetry-breaking}

\newcommand{\rem}[1]{}

\newcommand{\drawsquare}[2]{\hbox{%
\rule{#2pt}{#1pt}\hskip-#2pt
\rule{#1pt}{#2pt}\hskip-#1pt
\rule[#1pt]{#1pt}{#2pt}}\rule[#1pt]{#2pt}{#2pt}\hskip-#2pt
\rule{#2pt}{#1pt}}

\newcommand{\Yfund}{\raisebox{-.5pt}{\drawsquare{6.5}{0.4}}}
\newcommand{\Yasymm}{\raisebox{-3.5pt}{\drawsquare{6.5}{0.4}}\hskip-6.9pt%
        \raisebox{3pt}{\drawsquare{6.5}{0.4}}}
\newcommand{\bYfund}{\overline{\Yfund}}
\newcommand{\bYasymm}{\overline{\Yasymm}}

\begin{document}
\tightenlines
\pagestyle{empty}

\preprint{
\begin{minipage}[t]{3in}
\begin{flushright}
UW-PT/98-08 \\
IASSNS--HEP--98/53\\
hep-ph/9806346 \\
\end{flushright}
\end{minipage}
}

\title{A One-Scale Model of Dynamical Supersymmetry Breaking}
\author{
Ann Nelson\thanks{Work supported in part by the
Department of Energy under Contract
DE--FG03--96ER40956.
}
}
\address{
Department of Physics, University of Washington\\
Seattle, WA 98195, USA
}
\author{
Matthew J. Strassler
\thanks{Work supported in part by National Science Foundation grant
NSF PHY-9513835 and by the W.M.~Keck Foundation.}}
\address{School of Natural Sciences,
Institute for Advanced Studies\\
 Princeton, NJ 08540, USA\\
{\tt strasslr@ias.edu}}

\maketitle
\begin{abstract}
A model of gauge-mediated supersymmetry breaking is constructed in
which the low-energy physics depends on a single dynamical scale.
Strong coupling dynamics of gauge theories plays an important role, in
particular through its effects on beta functions and through
confinement. The model does not have distinct messenger and
supersymmetry-breaking sectors.  The scale of supersymmetry breaking
is of order 10-100 \TeV, implying that the decay of the
next-to-lightest superpartner into the gravitino is
prompt. Superoblique corrections are enhanced.  A Dirac fermion and
one complex scalar, in a $10$ or $\overline{10}$ of (global) $SU(5)$,
are predicted to be relatively light and to satisfy certain mass
relations with the standard model squarks and sleptons.
\end{abstract}


\newpage
\pagestyle{plain}
\narrowtext






\section{Introduction}
\label{sec:Intro}

Supersymmetry, broken dynamically, solves the gauge hierarchy problem
\cite{witten}. Communicating supersymmetry breaking to the
superpartners of the minimal supersymmetric standard model (MSSM) via
the ordinary \smt\ gauge interactions provides a natural explanation
for degeneracy of the squarks and sleptons, avoiding large
contributions to quark and lepton flavor violation from the
superpartners. Such gauge-mediated supersymmetry breaking (GMSB)
models
\cite{oldGMSBa,oldGMSBb,oldGMSBc,oldGMSBd,oldGMSBe,newGMSBa,newGMSBb,newGMSBc}
are therefore very attractive and have received much attention
recently
\cite{msa,msb,msc,msd,mse,msf,msg,msh,msi,msj,msk,msl,msm,msn,mso}.
In principle, with GMSB, the scale of supersymmetry breaking $\F$
could be as low as $\sim 4\pi m_W/\alpha_W\sim 30$ TeV, where $m_W$ is
the weak scale and $\alpha_W$ is the weak fine structure
constant. With R parity conservation and with $\F\ltap 1000$~TeV,
there is the exciting prospect of observing the decay of the
next-to-lightest superpartner (NLSP) into the gravitino $\tilde G$ and
ordinary particles in a typical particle physics detector
\cite{msa,msb,msc,msd,mse,msf,msg,msh,msi,msj,msk,msl,msm,msn,mso}.
Furthermore, the decay rate into $\G$ scales as $F^2$ and gives a
sensitive probe of the supersymmetry breaking sector.  One candidate
for such an event, with two leptons, two photons, and missing energy,
has already been reported by the CDF collaboration \cite{theevent}.
Interpreting the photons as coming from the prompt decay of the
lightest neutralino into a photon and gravitino requires $\F\ltap 100$
TeV.

Nearly all explicit models of dynamical supersymmetry breaking with
GMSB predict a supersymmetry breaking scale $\F$ which is in the range
$10^3$--$10^{8}$~TeV --- too high to allow the prompt decay of the
NLSP into the gravitino
\cite{newGMSBa,newGMSBb,newGMSBc,sima,simb,simc,simd,sime,simf,simg,simh},
\cite{nmsa,nmsb,nmsc,nmsd,nmse,nmsf,nmsg,nmsh}.  (The only reported
exceptions \cite{stmsa,stmsb,stmsc} may or may not have a
strongly-coupled local minimum with broken supersymmetry, and also
have supersymmetric vacua.)  There are several reasons why GMSB models
typically have such a large supersymmetry breaking scale.  In all GMSB
models, the ordinary superpartners gain mass though loops involving
``messenger'' particles which carry \smt\ quantum numbers and have a
nonsupersymmetric spectrum. In some models $\F$ must be high because
supersymmetry breaking is communicated to the messengers via some weak
coupling from a dynamical supersymmetry breaking (DSB) sector
\cite{newGMSBa,newGMSBb,newGMSBc,sima,simb,simc,simd,sime,simf,simg,simh}.
In other, more aesthetically pleasing, models the messengers are an
integral part of the DSB sector
\cite{nmsa,nmsb,nmsc,nmsd,nmse,nmsf,nmsg,nmsh,nmsi}, as first
suggested by Affleck, Dine, and Seiberg \cite{ADSd}. The latter models
typically have many particles carrying \smt\ interactions, with
perturbative unification of the standard model gauge couplings only
possible if the additional particles are very heavy. Most such models
constructed to date achieve \smt\ unification by having two or more
scales involved in the \ssb\ dynamics, with the majority of the new
particles at a messenger mass scale $M$ which is much heavier than the
supersymmetry breaking scale.  Since the ordinary superpartner masses
are proportional to $F/M$, both scales are required to be rather high.

In this paper we present a model of dynamical supersymmetry breaking
whose low-energy physics is determined by a single energy scale of
order $100$ TeV. To our knowledge, this is the first explicit, natural
model with no supersymmetric minima, all scales generated dynamically,
and prompt decay of the NLSP into the gravitino. The model has no
segregation of DSB and messenger sectors, is completely chiral and
contains no fundamental gauge singlets.  It has limits in which one
can show that the global minimum of the potential breaks supersymmetry
but leaves color and electromagnetism unbroken.  Perturbative
unification of the \smt\ interactions is possible and gives the usual
successful prediction for $\sin^2\theta_W$.  As our example is
strongly coupled at the supersymmetry breaking scale, it is somewhat
less predictive than most explicit GMSB models. Still, many of the
usual GMSB predictions survive. Unfortunately it is difficult to solve
the $\mu$ problem in this model.

 In \SEC{sketch} we give an quick overview of the model.  We prove the
model breaks \susy\ in \SEC{susybreaks}.  Since the model has
complicated behavior due to strong coupling, we review various facts
about strong dynamics in \none\ \susic\ gauge theories in
\SEC{betagammas}.  In section \SEC{careful}, we give a more detailed
discussion of the model's dynamics, and then justify our claims
carefully.  The low-energy properties of the model are explained in
\SEC{phenomena}; the reader who is mainly interested in the
implications for experiment can skip to this section.  The conclusion
contains a summary of our results.

\section{The Model: A First Sketch}
\label{sec:sketch}

The model we consider contains, in addition to the standard model, a
dynamical \ssb\ sector with gauge group $Sp(4)\times SU(3)\times
SU(2)$. We will refer to the coupling of the standard model
color and weak interactions as $g_3^{SM}$ and $g_2^{SM}$ to
distinguish them from $g_3$ and $g_2$ of the 3-2 \ssb\ sector. The
matter content of the model is given in Table 1.  The $SU(5)$ in the
last column is the usual grand unification group containing the
standard model.  Although we do not require gauge group unification
and treat $SU(5)$ merely as an approximate global symmetry, we
consider only complete multiplets of $SU(5)$ in order that standard
model gauge coupling unification be maintained.  (The $SU(5)$
assignments could be charge conjugated without changing the model.)

\bigskip

\begin{tabular}{|l|c|c|c|c|}
\hline
\hfil  &\hfil Sp(4) \hfil &\hfil SU(3)\hfil &\hfil SU(2) \hfil&\hfil
SU(5)$_G$ \hfil  \\ \hline
&&&&
\\ [-8pt]
 $q$       &1        &$\Yfund$   &$\Yfund$   &1        \\ \blanklinefour
 $\bar u$  &1        &$\bYfund$  &1          &1        \\ \blanklinefour
 $\bar d$  &1        &$\bYfund$  &1          &1        \\ \blanklinefour
 $\ell$    &1        &1          &$\Yfund$   &1        \\ \blanklinefour
 \blanklinefour
 $\bar T$  &$\Yfund$ &$\bYfund$  &1          &1        \\ \blanklinefour
 $\bar V$  &$\Yfund$ &1          &1          &$\bYfund$  \\ \blanklinefour
 \blanklinefour
 $A$       &1        &1          &1          &$\Yasymm$        \\
\blanklinefour
 $B$       &1        &$\Yfund$   &1          & $\Yfund$        \\
\blanklinefour
 $C$       &1        &$\bYfund$  &1          &1        \\
\hline
\end{tabular}
\vskip .2 in Table 1.
{\it Quantum numbers of chiral superfields in the model.  $SU(5)_G$ is
a global symmetry containing the standard model.  }
\vskip .2 in

\bigskip

First, we give a brief  motivation for the model.  The
fields $q,\bar u,\bar d,\ell$ make up the matter content of the famous
supersymmetry-breaking $SU(3)\times SU(2)$ model of Affleck, Dine and
Seiberg \cite{ADSa,ADSb,ADSc} (the ``3-2 model'').  The $Sp(4)$ gauge
group has a total of eight fields in its fundamental representation,
coming from $\bar T$ and $\bar V$, and consequently will
confine at low energies \cite{kippsp}, at a scale
$\Lc$. The resulting massless bound states $\bar A=(\bar V\bar V)$,
$\bar B=(\bar T\bar V)$, $\bar C=(\bar T\bar T)$, have the correct
quantum numbers to pair up with the fields $A,B,C$ and become massive.
Thus, below the $Sp(4)$ confining scale the theory will consist of the
standard model, the massless fields of the 3-2 model, and massive
fields which couple to both sectors and act as messenger fields by
communicating the supersymmetry breaking of the 3-2 model to the
standard model.

For the model to behave in this way requires a superpotential
\be{fullsup}
W = W_{SM} + W_{3-2} + W_{m} + W_{s}
\ee
where $W_{SM}$ is the standard model superpotential,
\be{ssbsup}
W_{3-2} = \lambda_0 q \bar u \ell
\ee
 is the usual superpotential of the 3-2 model, and
\be{ABCsup}
W_{m} = y_A A\bar V\bar V+y_B  B\bar T\bar V+y_C C\bar T\bar T
\ee
serves to give masses to the messenger fields.
Finally, the couplings
\be{decaysup}
W_{s} = h_1 Cq\ell + h_2 C\bar u\bar d
\ee
are not needed to ensure an acceptable pattern of symmetry breaking
but will help avoid having stable heavy messenger particles.  Note
that this is the most general renormalizable superpotential consistent
with gauge symmetry which does not couple the MSSM to the DSB sector.
For simplicity in the following discussion, we will assume the
$h_{1,2}$ couplings are small and have little effect on the dynamics,
although this is not essential or even likely, since they get enhanced
by strong coupling effects.

The dynamics of the theory is intricate. The $SU(3)$ gauge coupling is
expected to flow slowly due to higher loop effects, and
approach a fixed point at extreme low energy.  As a result, the scale
$\Lambda_3$ is washed out by the dynamics.  By taking
$\Lambda_4\ll\Lambda_3$, we can arrange that $Sp(4)$ confinement, at
the scale $\Lc$, occurs when the coupling $g_3$ is
substantial.  Associated masses of order $\Lc$ for the $B,\bar B$
and $C,\bar C$ fields remove all $SU(3)$-charged fields except those
of the 3-2 model.  The $SU(3)$ beta function then becomes large,
causing $g_3$ immediately to blow up, breaking \susy.  We therefore
expect the scale $\F$ of \susy\ breaking to be of order the dynamical
scale $\Lc$.

Thus, in this model the messengers and the
\susy\ breaking lie at or near the same scale, which we take to be of
order 10--100 TeV.  Note that the model has neither vectorlike matter nor
non-dynamical mass scales.  The gravitino is light, and its properties
are similar to those of other low-scale GMSB models; it can serve to
explain the $e^+e^-\gamma\gamma$ event observed at CDF
\cite{theevent}.

The model has another feature which appears in certain regions of
parameter space.  As we will see below, the fact that the field $A$ is
a 4-3-2 gauge singlet tends to make the $A\bar A$ dynamical mass
smaller than $\Lc$.  As a result, the
Dirac fermion $\psi_A,\psi_{\bar A}$ and the complex scalar $A$ might
(but need not) be much lighter than $\Lc$.  (The scalar $\bar A$ is a
composite of strongly interacting fields and will get a large \ssb\
mass.)  The mass spectrum of these fields is interesting
and will be discussed in detail in \SSEC{BBar}.

The effects on the standard model superpartners resemble those in
usual GMSB models that have heavy messengers charged in both the \ssb\
and standard model groups. However, because the \ssb\ sector is
strongly interacting, and because the messengers have masses near
$\Lc$, there is no separation of scales in this model.  In fact one
cannot really talk of a ``messenger sector''; the strong dynamics as a
whole is responsible for the message.  The effective
action below the scale $\Lc$ is already far from supersymmetric.  This
can make some aspects of the model quite different from GMSB models
with weakly-coupled messenger sectors.  For example, the overall scale
of the gaugino masses is unrelated to that of the sfermion masses
because of the strong dynamics at the scale $\Lc$.  Also, there are
relatively large ``superoblique'' corrections to gaugino couplings,
of order $\alpha^{SM}_i \log\left(\Lc/m_{\psi_A}\right)/(4 \pi)$.
Still, the strong couplings of the model preserve an approximate $SU(5)$
global symmetry, which ensures that masses of different gauginos are
related by standard model gauge couplings, and similarly for sfermion
masses.

\section{Breaking of Supersymmetry}
\label{sec:susybreaks}

  In this section we will demonstrate that the model breaks
supersymmetry, first showing the model has no flat directions at the
quantum level, and then demonstrating that supersymmetry is broken for
a generic choice of parameters.

\subsection{Absence of Flat Directions}
\label{ssec:noflat}

Our model has no flat directions at the quantum mechanical level, and
hence no supersymmetric minima infinitely far away in field
space. Here we study the classical flat directions, which are labelled
by holomorphic invariants built from the chiral fields. To simplify
the discussion, we rescale all Yukawa couplings to one.

Any
holomorphic invariant involving fields charged under $Sp(4)$ must
involve one of $\bar V\bar V$, $\bar T\bar V$ or $\bar T\bar T$.  The
first two are set to zero by the F-flatness conditions $\partial
W/\partial A = 0$ and $\partial W/\partial B = 0$, while $\partial
W/\partial C = 0$ assures $\bar T\bar T=-(q\ell+\bar u\bar d)$.  Using
$\partial W/\partial \bar u=0= \partial W/\partial \bar d$ and
antisymmetry, one can show the operators $\bar T\bar T C,\bar T\bar
T\bar u,\bar T\bar T\bar d$ are all zero. The operator $\bar Tqq\bar
T$ also vanishes; since $qqql$ is identically zero, the F-flatness
conditions $\partial W/\partial C = 0 = \partial W/\partial \ell$
imply that $\bar Tqq\bar T\propto \bar uqq\bar d\propto Cqq\bar d$,
which in turn is zero by $\partial W/\partial \bar u=0$. All operators
which involve only the 3-2 fields $q,\bar u,\bar d$ and $\ell$ must be
zero.  Finally there are some classical flat directions which combine
$A$, $B$ and $C$ with the 3-2 fields $q,\bar u,\bar d,\ell$.  However,
as we now show, even these are removed by quantum mechanical effects.

Along any classically flat direction with expectation values for $A$,
$B$ or $C$, some of the fields in the fundamental representation of
$Sp(4)$ (components of $\bar V$ and/or $\bar T$) will be massive.  The
number of remaining massless fundamentals may be six, four, two or
zero.  In each case, strong-coupling dynamics of the $Sp(4)$ group
\cite{kippsp} then generates a potential energy.  If the number
remaining is six, then the classical moduli space of the $Sp(4)$
theory is modified by the constraint that $V^5T^3 \sim \Lambda_L^8$
(here $\Lambda_L$ is the {\it low-energy} $Sp(4)$ strong-coupling
scale.)  The requirements $\partial W/\partial A = \partial W/\partial
B = \partial W/\partial C = 0$ imply that $V^5T^3 =0$ for a
zero-energy vacuum, in contradiction to the previous condition.  If
the number remaining is four (two), the $Sp(4)$ theory generates an
Affleck-Dine-Seiberg superpotential via instantons (gaugino
condensation) which again lifts the flat directions.  And if all of
the fields $\bar V$ and $\bar T$ are massive, then gaugino
condensation generates a superpotential $W=\Lambda_L^3$, where again
$\Lambda_L$ is the {\it low-energy} $Sp(4)$ strong-coupling scale,
related by $\Lambda_L^9 = A^2BC \Lambda_4^5$ to the high-energy
$Sp(4)$ strong-coupling scale $\Lambda_4$.  Thus, in terms of the
original fields, the low-energy superpotential is $W = ( A^2BC
\Lambda_4^5)^{1/3}$, and the equations $\partial W/\partial A =
\partial W/\partial B = \partial W/\partial C = 0$ then require the
expectation values of $A,B,C$ to vanish for a zero-energy vacuum.

\subsection{Dynamical Supersymmetry Breaking}
\label{ssec:dsb}

Having established that there are no flat directions in the model, we
now proceed to show that the model breaks supersymmetry.  We need only
show this in a particular range of $\Lambda_4$, $\Lambda_3$ and
$\Lambda_2$ (these are the strong coupling scales for the three new
gauge groups.)  Holomorphy in $\Lambda_i/\Lambda_j$ ensures there will
be no phase transitions as these couplings are varied; at worst there
may be singular points for special values of $\Lambda_i/\Lambda_j$,
which we can choose to avoid.  Thus, if supersymmetry is broken for
any open set of values for $\Lambda_i/\Lambda_j$, then it will
be broken for most values of $\Lambda_i/\Lambda_j$.

Although we will eventually construct a model in which
$\Lambda_3>\Lambda_4\gg\Lambda_2$, it is easiest to show supersymmetry
is broken in the regime $\Lambda_4\gg\Lambda_3\gg\Lambda_2$.  The
large separation of scales allows us to treat the strong dynamics of
the gauge groups one at a time, with the remaining weakly-coupled
groups (including the standard model) serving as spectators.  First,
the $Sp(4)$ gauge group becomes strongly coupled at the scale
$\Lambda_4$.  It confines, and the low-energy dynamics is given in
terms of the mesons $\bar A=(\bar V\bar V)/\Lambda_4$, $\bar B=(\bar
T\bar V)/\Lambda_4$, $\bar C=(\bar T\bar T)/\Lambda_4$, which are
massless in the absence of a tree-level superpotential.  Note that we
have normalized the mesons to have dimension one, as is appropriate at
low energy.  The resulting matter content (aside from the standard
model fields) is given in Table 2.

\bigskip

\begin{tabular}{|l|c|c|c|c|}
\hline
\hfil &\hfil SU(3)\hfil &\hfil SU(2) \hfil&\hfil
SU(5)$_G$ \hfil  \\ \hline
&&&
\\ [-8pt]
 $q$       &$\Yfund$   &$\Yfund$   &1        \\ \blanklinethree
 $\bar u$  &$\bYfund$  &1          &1        \\ \blanklinethree
 $\bar d$   &$\bYfund$  &1          &1        \\ \blanklinethree
 $\ell$    &1          &$\Yfund$   &1        \\& & &  \\
 \blanklinethree
 $\bar A$        &1          &1          &$\bYasymm$     \\ \blanklinethree
 $\bar B$        &$\bYfund$   &1          & $\bYfund$    \\ \blanklinethree
 $\bar C$         &$\Yfund$  &1          &1        \\
 \blanklinethree
 $A$        &1          &1          &$\Yasymm$        \\ \blanklinethree
 $B$        &$\Yfund$   &1          & $\Yfund$        \\ \blanklinethree
 $C$         &$\bYfund$  &1          &1        \\
\hline
\end{tabular}
\vskip .2 in
Table 2. {\it Quantum numbers of chiral superfields after $Sp(4)$ confines.}
\vskip .2 in

\bigskip

The strong dynamics of the theory generates a dynamical superpotential
\cite{kippsp}
\be{dynsup}
W_{dyn} = {\bar A\bar A\bar B\bar C \over \Lambda_4} \ .
\ee
The superpotential of the theory is now
\be{fullsupII}
W = W_{SM} + W_{3-2} +
W_{M} + W_{s} + W_{dyn}
\ee
where $W_{SM}$, $W_{3-2}$ and $W_s$ are
the same as before and
\be{ABCsupII}
W_M = y_A\Lambda_4 A\bar A+y_B\Lambda_4
B\bar B+y_C\Lambda_4 C\bar C
\ee
is $W_m$ reexpressed in terms of the composite fields.  This last set
of interactions results in masses of order $\Lambda_4$ for the
fundamental fields $A$, $B$ and $C$ and the composite mesons $\bar A$,
$\bar B$ and $\bar C$.  Without changing the infared dynamics, we may
integrate out the massive fields, eliminating $W_M$ and $W_s$ from the
superpotential and changing the K\"ahler potential by high dimension
operators.  The 3-2 sector and standard model sector are then
connected only by irrelevant interactions, and the former, as shown by
Affleck, Dine and Seiberg, breaks supersymmetry at a scale determined
by $\lambda_0$, $\Lambda_2$ and $\Lambda_3$.

\section{Beta functions and anomalous dimensions}
\label{sec:betagammas}

Our model exhibits a number of interesting and subtle strong-coupling
phenomena, which we will discuss carefully.  Because of this, we begin
with a review of some dynamical relations in supersymmetric theories,
which provide tools for semi-quantitative analysis of strongly-coupled
theories.  These tools will not be powerful enough to make our results
unambiguous, but they will provide evidence that the model exhibits
the qualitative features which we need to make use of.

In \none\ \susic\ theories there are relationships between beta
functions and anomalous dimensions.  A Yukawa coupling
$y_0$ in the superpotential $W_Y = y_0\phi_1\phi_2\phi_3$ has a beta
function which is a function of all the other Yukawa couplings $y_i$
and gauge couplings $g_k$ in the theory.
Non-renormalization theorems in \none\ \susic\ theories  ensure
that all vertex functions are trivial and that all running of couplings
comes through wave function renormalization.
Consequently, the beta function of $y_0$ is related in a
simple way to the anomalous mass dimensions $\gamma_n(y_0,y_i,g_k)$
of the fields $\phi_n$.
\be{ybeta}
\beta_{y_0} = {1\over 2} y_0
\left[\gamma_1(y_0,y_i,g_j)+\gamma_2(y_0,y_i,g_j)+
\gamma_3(y_0,y_i,g_j)\right]
\ee
The running of gauge couplings is slightly more complicated, but
still related linearly to the anomalous dimensions of the fields.
According to \cite{SVa,SVb} the coupling $g_k$ runs as
\be{SVbeta}
\beta_{g_k} = -{g_k^3\over 16 \pi^2}
{3C_2(G_k)-\sum_p T(\phi_p)[1-\gamma_{\phi_p}]\over
1-{g_3^2\over 8\pi^2}C_2(G_k)} \ .
\ee
Here $C_2(G_k)$ is the second Casimir operator of the gauge group $G_k$ for
which $g_k$ is the coupling, the sum in the numerator is over all
matter fields $\phi_p$, $T(\phi_p)$ is half the index of the
representation of $\phi_p$ under $G_k$, and $\gamma_{\phi_p}$ is the
anomalous dimension of $\phi_p$.  Note that to leading order in the
couplings this expression gives
\be{oneloop}
\beta_{g_k} = -{g_k^3\over 16 \pi^2}b_0 \ ; \ b_0\equiv
{3C_2(G_k)-\sum_p T(\phi_p)} \ .
\ee
where $b_0$ is the well-known coefficient of the one-loop correction
to the gauge coupling.

There are also conditions which follow from the \none\ superconformal
algebra, which constrains the properties of theories at exact or
approximate fixed points.  One of these is the ``unitarity
condition''.  In any four-dimensional conformal field theory,
unitarity implies that no gauge invariant operator (except the unit
operator) can have dimension less than one \cite{cfta,cftb,cftc}.  Any
operator whose dimension is exactly one must satisfy the Klein-Gordon
equation, and cannot interact with any other fields.  These facts will
apply also to any operator which is gauge variant only under a very
weakly-coupled gauge group: as it must have dimension greater than one
in the limit of zero gauge coupling, perturbation theory in the small
gauge coupling ensures that its dimension cannot be much below one.  A
related consequence is that when all gauge couplings are small, a
field with a large Yukawa coupling always has a positive anomalous
dimension.

Another condition relates the R charge of a chiral operator and its
anomalous dimension \cite{cfta,cftb,cftc}.  At a superconformal fixed point
there is a special R current that appears in the same superconformal
multiplet as the energy-momentum tensor and the supersymmetry current.
At a fixed point, the dimension of a chiral operator is ${3\over 2}$
times its R charge, from which its anomalous dimension may be
calculated.  (The result always agrees with results which follow from
the beta functions above.)  An important implication of this result is
that, since R charges are additive, the dimension of a composite
chiral operator is equal to the sum of the dimensions of its chiral
constituents.  This can be restated as resulting from the absence of
short-distance singularities when any two chiral operators are brought
to the same point.  (Similar statements of course apply to antichiral
operators.)  Unfortunately, the general theory has an infinite set of R
symmetries, and often it is impossible to determine which of them
appears in the multiplet of currents.

An important corrolary of the above results involves the running of
Yukawa couplings.  Consider a set of fields with gauge couplings $g$
and Yukawa couplings $y$ with anomalous dimensions $\gamma(y,g)$.
Unitarity ensures that $\gamma(y,0)$ is positive.  It is easy to check
that $\gamma(0,g)$ is negative for a charged field for small $g$.
Together with \Eref{ybeta}, these imply that a Yukawa coupling
involving charged fields will be irrelevant for $g\ll y$ but may
become relevant as $g$ becomes of order $y$.  By contrast, a Yukawa
coupling for three gauge-neutral fields is always irrelevant.

\section{The Model: A Careful Rendering}
\label{sec:careful}

We now provide an detailed overview of the model, making claims
about the dynamics which we justify later in this and in the
following section.  Our approach is semi-quantitative and relies on the
dynamical relations described in \SEC{betagammas}.

We consider the model in the regime $\Lambda_3>\Lambda_4\gg\Lambda_2$.
The standard model couplings and $g_2$ are smaller than the gauge
couplings $g_3,g_4$, and can be neglected in most of our analysis.
The Yukawa couplings $\lambda_0,y_A,y_B,y_C$ may not be small (though
we do assume for simplicity that $h_{1},h_2$ are small.)  The
renormalization-group flow of the gauge and Yukawa couplings cannot be
known exactly, but can be analyzed using \SEC{betagammas}.  A possible
form for the behavior of the couplings is sketched in Figure 1.

\begin{figure}
\centering
\epsfxsize=4in
\hspace*{0in}
\epsffile{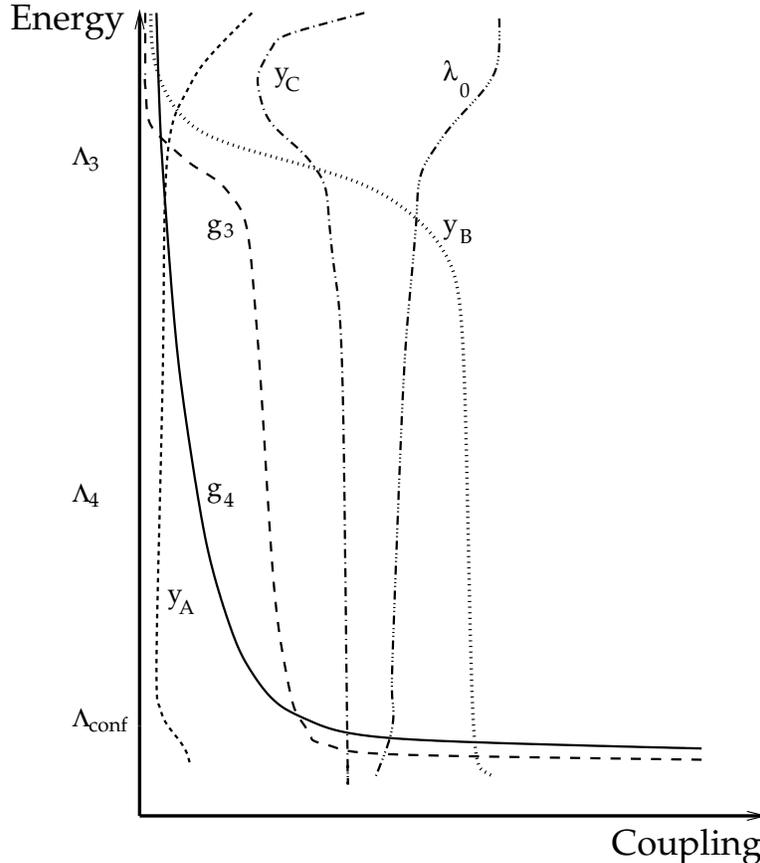}
\caption{Possible renormalization group flow for the most
important couplings.  The actual flow could be very different;
this figure is for illustration only.}
\label{fig:flow}
\end{figure}

Before any gauge couplings are strong (assuming there is such a range)
the Yukawa couplings are all irrelevant.  Once $g_3$ becomes strong,
however, $\lambda_0$, $y_B$ and $y_C$ may become relevant; they are
certainly relevant when they are small, and therefore they become or
remain large.  By contrast, $y_A$ is irrelevant and may become small.
A theory of $SU(3)$ with seven triplets and antitriplets, along with
some $SU(3)$ singlets and Yukawa couplings, is expected to flow to a
conformal field theory in the far infrared (see \SSEC{suthree}.)  The
gauge coupling $g_3$ and the relevant Yukawa couplings will run slowly
as they gradually approach an infrared fixed point. Our knowledge of
the properties of this fixed point is limited.  We know that it should
preserve the global $SU(5)$ symmetry. We also know that it occurs
outside of perturbation theory, and so the theory begins to look
conformal only at scales far below $\Lambda_3$.  In fact the theory is
unlikely to be extremely close to the fixed point when supersymmetry
breaks, although $SU(5)$ will still be a good approximate symmetry
(see \SSEC{sufive}.)

The $Sp(4)$ group has a negative beta function when $g_3$ is small.
The strong coupling effects of $g_3$ might in principle change the
sign of $\beta_{g_4}$, but they do not, as shown in \SSEC{spfourbeta}.
Consequently, the $Sp(4)$ coupling grows.  As it becomes strong the
coupling $y_A$ will become relevant, though it may not have a large
energy range in which to grow.  The other relevant Yukawa couplings
are expected to remain large.  The coupling $g_4$ is not expected to
reach a fixed point (a conspiracy would be needed to allow for this
possibility, see \SSEC{spfourbeta}) and so at some scale
$\Lambda_{conf}$ --- a real physical scale, not to be confused with
the holomorphic scale $\Lambda_4$ --- $Sp(4)$ confines.  Below this
scale, meson degrees of freedom $\bar A=\bar V\bar V/\Lambda_4, \bar
B=\bar V\bar T/\Lambda_4, \bar C=\bar T\bar T/\Lambda_4$ best describe
the long distance physics, and the theory possesses the fields of
Table 2.  The factors of $\Lambda_4$ are convenient in order to have a
holomorphic definition of these fields which has mass dimension one.
However, an additional nonholomorphic dimensionless factor is
necessary for these composite fields to be canonically
normalized. While this factor cannot be computed, it can be estimated
on physical grounds to be such that if $y_A$ is large (of order
$4\pi$), then the mass of $A$ is of order $\Lc$.  Similar
considerations normalize $B$ and $C$.\footnote{Note that the
superpotential \eref{ABCsupII} is misleading; the masses of $A$, $B$
and $C$ are of order $\Lc$, not $\Lambda_4$, as a result of this
normalization factor, which appears in the K\"ahler potential.}

 Since the couplings $y_B$ and $y_C$ are strong, the fields $B,\bar B$
and $C,\bar C$ have masses of order $\Lc$.  This leaves the $SU(3)$
group below this scale with only two triplets and a large beta
function; $g_3$ blows up at once, at a scale of order $\Lc$.  Since
$\lambda_0$ is large, \susy\ is broken immediately by $SU(3)$ strong
coupling effects, with $\F$ close to $\Lc$ (see \SSEC{blowup}.)\footnote{In
an entirely strongly coupled theory, with no small dimensionless
parameters, naive dimensional analysis \cite{ndaa,ndab} would give $F\sim
\Lc^2/(4\pi)$. Since in our theory the weak coupling $g_2$ can
suppress supersymmetry breaking, as shown in \SSEC{dsbvac}, we will keep $F$
a free parameter.}  This vacuum is likely to preserve color and
electromagnetism, as argued in \SSEC{dsbvac}.

The Yukawa coupling $y_A$ may be driven small, as explained in
\SSEC{lightmess}, and the fields $A,\bar A$ therefore may have a
supersymmetric mass somewhat smaller than $\Lc$.  There are also light
particles from the 3-2 sector: the goldstino and one other particle
whose presence is required by anomaly matching \cite{threetwo}.
Finally, there are the light fields of the standard model.  We now
would like to integrate out the dynamics above the scale $\Lc$ and
write an effective theory for the light degrees of freedom valid below
this scale.  However, the dynamics of the supersymmetry breaking is
strongly coupled, and reliable quantitative analysis is impossible.  A
qualitative approach to this effective lagrangian is therefore
necessary. We will use naive dimensional analysis to estimate various
quantities which cannot be computed \cite{ndaa,ndab}.  Although there is no
empirical evidence that this works for strongly coupled theories other
than QCD, we expect such estimates to be off by no more than an order
of magnitude.

The act of integrating out the \ssb\ sector introduces soft \ssb\
terms at the scale $\Lc$.  Fields which couple strongly to the \susy\
breaking, such as the composite scalars $\bar A$, will have \ssb\
masses of order $4\pi F/\Lc$ (see \SSEC{AAbar}.) The $\bar A, A$
fermions obtain supersymmetric masses of order $y_A\Lc/(4\pi)$ ---
chiral symmetry protects them against \ssb\ masses.  The $A$ scalars
discover supersymmetry breaking only via the weak couplings $y_A$ and
the standard model gauge couplings $\alpha^{SM}_k$, and hence (see
\SSEC{BBar}) have masses of order Max[$y_A\Lc/(4\pi), \alpha^{SM}_k
F/\Lc $].  All the MSSM fields couple to the \ssb\ sector via
standard model gauge couplings, as in usual GMSB models.  A modified
version of the usual results \cite{newGMSBa,newGMSBb,newGMSBc} applies
--- both gauginos and scalars acquire masses of order $ \alpha^{SM}_k
F/ \Lc$ (see \SSEC{BBar}.)  Since $\F$ lies close to $\Lc$, we take
$\Lc\sim 30-1000$ TeV and $\F\sim 10-100$ TeV.  Such low values of
$\F$ lead to rapid decays of the next-to-lightest superpartner to a
gravitino, as discussed in
\cite{msa,msb,msc,msd,mse,msf,msg,msh,msi,msj,msk,msl,msm,msn,mso}.

The large splittings in the messenger multiplets leads to substantial
superoblique corrections \cite{superobliquea,superobliqueb}; these are
estimated in \SSEC{BBar}.  Trilinear scalar couplings and the $\mu$
problem --- a difficulty in this model --- are discussed in
\SSEC{muterm}.  Finally, in \SSEC{messdecay}, we explain a couple of
ways to avoid stable messenger fields, as required to prevent conflict
with experiment.

\subsection{The $SU(3)$ physics}
\label{ssec:suthree}

It has been argued that the theory of $SU(3)$ with seven triplets and
antitriplets reaches an infrared fixed point \cite{NAD}.  At this
fixed point the charged fields have negative anomalous dimensions, so
Yukawa couplings involving either two charged fields and a singlet or
three charged fields (such as $\lambda_0,y_B,y_C$) are relevant and
drive the theory away from this fixed point.  Does the theory flow to
another fixed point in the infared, or behave altogether differently?
It is possible to construct theories, with relevant Yukawa couplings,
which preserve no R charges that could be part of a unitary conformal
field theory; in such cases a low-energy fixed point is unlikely
(though not impossible, since accidental R symmetries may arise in the
infrared.)  However, in our theory, there are R charges, and
corresponding candidate infrared fixed points, which are consistent
with unitarity and with the Yukawa couplings of the superpotential of
eq.~\eref{fullsup}. These would preserve the $SU(5)$ global symmetry
which contains the standard model.  On the other hand, there is
insufficient flavor symmetry to determine the R charges at the fixed
point and confirm that unitarity is not violated.  Thus we cannot
provide an argument that the couplings we have chosen do or do not
lead to a fixed point.

However, for our present purposes, such an argument is not really
necessary.  Even if the $SU(3)$ theory does not reach an approximate
fixed point, it is likely that its beta function will be very small,
much smaller than the one-loop estimate.  (This is certainly true if
the Yukawa couplings are small, due to $SU(3)$ two-loop corrections.)
A slow-running coupling constant tends to wash out physical effects
involving the scale $\Lambda_3$. Meanwhile, in the effective theory
below the scale of $Sp(4)$ confinement, the $SU(3)$ beta function is
rather large, so for a substantial range of $\Lambda_3$ and $
\Lambda_4$, the physical scales associated with strong $Sp(4)$ and
$SU(3)$ dynamics are close together, and close to the supersymmetry
breaking scale.  Furthermore, if $SU(5)$ is a roughly approximate
symmetry at high energy, it will be preserved even when the $SU(3)$
coupling becomes large.  We will explain how this occurs in
\SSEC{sufive}.

\subsection{The $Sp(4)$ beta function}
\label{ssec:spfourbeta}

We need to show that $\beta_{g_4}<0$ at all scales, so that $Sp(4)$
confines rather than reaching a conformal fixed point.  When $g_4$ is
small, this can be proven.  When $g_4$ is large, a proof is not
possible, but we will show it is unlikely that $\beta_{g_4}$ reaches a
zero.

The beta function for $Sp(4)$ is given by
\be{spbeta}
\beta_{g_4}= -{g_4^3\over 16 \pi^2}
{9-{3\over 2}[1-\gamma_{\bar T}]-{5\over 2}[1-\gamma_{\bar V}]\over
1-{g_4^2\over 2\pi^2}}
= -{g_4^3\over 16 \pi^2}
{5+{3\over 2}\gamma_{\bar T}+{5\over 2}\gamma_{\bar V}\over
1-{g_4^2\over 2\pi^2}} \ .
\ee
Above $\Lambda_3$ the one-loop formula is approximately correct
\be{sponeloop}
\beta_{g_4}= -5{g_4^3\over 16 \pi^2}\ .
\ee
and $g_4$ grows logarithmically with coefficient $5$.  However, once
$g_3$ is large we expect that $\bar T$ has a negative anomalous
dimension, and once $y_B$ is large then $\bar V$ will have a positive
anomalous dimension (by unitarity).  The effect of $\gamma_{\bar T}$
will tend to make $g_4$ run more slowly, while that of $\gamma_{\bar V}$
will tend to make it run more quickly.

The operator $\bar T\bar T\bar T$ is charged only under $Sp(4)$.  When
$g_4\ll 1$, the unitarity condition demands that $\bar T$ have
dimension close to or greater than $1/3$, and thus $\gamma_{\bar
T}\geq -4/3 - \CO(g_4^2)$.  It follows that the function $5+{3\over
2}\gamma_{\bar T}+{5\over 2}\gamma_{\bar V}$, which appears in the
numerator of the beta function, is greater than or of order 3 in this
regime.  Thus $g_4$ remains asymptotically free, but may run more
slowly than the one-loop estimate, leading $\Lc$ to be much lower than
$\Lambda_4$.

For large $g_4$ a different argument is necessary.  Once $g_4$ becomes
large, then, in the limit $g_2=g_i^{SM}=0$, $\bar V\bar V$ and $\bar
V\bar T\bar T\bar T$ are gauge invariant, implying $\gamma_{\bar
V}>-1$, $3\gamma_{\bar T}+5\gamma_V>-10$.  Unfortunately this allows
the $Sp(4)$ beta function to be arbitrarily close to zero, and it
could reach zero when the gauge couplings $g_2,g_i^{SM}$ are non-zero
or when other small $SU(5)$ violation is accounted for.  While we
cannot rule this out, we note that it requires extreme values of
$\gamma_{\bar T}$ which are unlikely to be attained.  It is more
likely that the $Sp(4)$ beta function remains negative, though rather
smaller in magnitude than normally expected for a theory undergoing
confinement. This suggests that $\Lc\ll\Lambda_4$, which is of
relevance in \SSEC{lightmess} below.

\subsection{Supersymmetry breaks at the scale $\Lc$}
\label{ssec:blowup}

Can we be sure that the $SU(3)$ coupling blows up close to the scale
$\Lc$ and not well below that scale?  Consider the case $\Lambda_4\gg
\Lambda_3$, with small $\lambda_0,y_B,y_C$.  In this case, the $Sp(4)$
theory confines at the scale $\Lambda_4$, leaving the theory with the
fields in Table 1.  The $SU(3)$ beta function coefficient $b_0$
changes from 2 to 3.  The fields $B,\bar B,C,\bar C$ have masses of
order $y_B\Lambda_4, y_C\Lambda_4$; below these scales, $b_0 = 7$.
The one-loop analysis for the $SU(3)$ coupling is reasonably good in
this case, and we find the coupling blows up at
\be{blowup}
\Lambda_{3L} =
\Lambda_4[y_B^5y_C]^{1/7}[\Lambda_4/\Lambda_3]^{2/7}.
\ee
The small fractional powers indicate that the $Sp(4)$ dynamics
controls the divergence of the $SU(3)$ coupling even when $g_3$ is
small at $\Lambda_4$.  We expect this will be all the more true when
$y_B,y_C$ are large and $\Lambda_3>\Lambda_4$, in which case the
$SU(3)$ coupling will be substantial, and slow-running, down to the
scale $\Lc.$

Since the strong $SU(3)$ physics drives supersymmetry breaking,
and since no couplings (except possibly $y_A$, see \SSEC{lightmess})
are small, we expect supersymmetry breaking to occur at
a mass scale $\F$ within an order of magnitude of $\Lc$.
We will show in \SSEC{dsbvac} that for sufficiently
small $g_2$ there is a \ssb\ vacuum with $F\sim g_2^{3/7}\Lc^2$
that preserves the standard model gauge group, and we will assume
the theory lies in this vacuum even for $g_2\sim 1$.

\subsection{An acceptable \ssb\ vacuum}
\label{ssec:dsbvac}

Although we have shown supersymmetry is broken, this is far from
showing that the vacuum of the theory is phenomenologically
acceptable.  In particular, the standard model gauge groups must not
be broken.

For $\Lambda_2$ sufficiently small and $\Lambda_4>\Lambda_3$, there is
a supersymmetry-breaking vacuum in which no field with standard model
charges gets an expectation value.  First, recall that when
$\Lambda_4\rarr\infty$ the low-energy theory is the 3-2 model with
massive messengers, and so when $\Lambda_2\rarr 0$, supersymmetry is
restored, even for finite $\Lambda_4$.  When $\Lambda_2=0$ there are
flat directions labelled by the operators $q\bar u,q\bar d,\ell$ which
carry only $SU(2)$ quantum numbers.  Classically, the superpotential
sets $q\bar u$ to zero, but $q\bar d$ and $\ell$ may still have
expectation values, both of which may be non-zero as long as $q\ell
\equiv q^i\ell^j\epsilon_{ij}=0$.  This allows the expectation values
\be{Dflat}
q = \left[\matrix{a &0&0\cr 0&0&0}\right] \ ; \ \bar d = \left[\bar a\
0\ 0\right] \ ; \ \ell= \left[\matrix {b \cr 0}\right]
\ee
where $|a|=|\bar a|.$ Along this flat direction the $Sp(4)$ gauge
group is unbroken while the $SU(3)$ gauge group breaks to $SU(2)_I$
with five massless flavors (the fields $q^1,\bar d$ are eaten and have
mass $|g_3 a|$, while $\bar u,\ell$ have mass $\lambda_0 b$; $B, \bar
T, C$ are massless.)  The strong coupling scale of this group is
$\Lambda_I=\Lambda_3^2 \lambda_0\vev{\ell/q\bar d}$. Take
$a,b,\Lambda_4>\Lambda_3 >\Lambda_I$; then the confining dynamics of
$Sp(4)$ gives mass to the messengers leaving the $SU(2)_I$ theory with
no massless flavors and a strong coupling scale $\Lambda_L^6 = y_B^5
y_C\lambda_0\Lambda_3^2 \Lambda_4^5 \vev{\ell/q\bar d}$.  The
low-energy superpotential is just given by the gaugino condensation in
these variables:
\be{Lsup}
 W_L = \left[{ y_B^5 y_C\lambda_0\Lambda_3^2
\Lambda_4^5 \vev{\ell}\over \vev{q\bar d}}\right]^{1/2}
\ee
Since $a,b\gg\Lambda_I$, the K\"ahler potential for these fields
is trivial, and so the potential energy along this direction is
\be{LV}
V_{I}(a,b)= |y_B^5 y_C\lambda_0\Lambda_3^2 \Lambda_4^5|
\left[{1\over 4 |ba^2|}+{|b|\over 2|a|^4} \right]
\ee
This is minimized at $a,b\rarr\infty$ with $|b/a^4|\rarr 0$.

We have assumed to this point that $\Lambda_2=0$.  For
$\Lambda_2\ll\Lambda_L$, the only effect of the gauged $SU(2)$ will be
the potential energy from the D-terms:
\be{DV}
V_{2}(a,b) = g_2^2
(|a^2|+|b^2|)^2 \ee
where $g_2$ is the $SU(2)$ coupling at energy
scales of order $a,b$.  The minimum of the potential $V_I+V_2$
can be seen to be
at $a\sim b\sim [g_2^{-2}y_B^5 y_C\lambda_0\Lambda_3^2
\Lambda_4^5]^{1/7} \gg \Lambda_3,\Lambda_4$, where the vacuum energy
density is of order
\be{MSdef}
F^2=g_2^{6/7}[y_B^5 y_C\lambda_0
\Lambda_3^2\Lambda_4^5]^{4/7}\ .
\ee
Notice that $a,b$ go to infinity and $F$ goes to zero as $g_2\rarr
0$, as expected; thus our assumption that $a,b\gg \Lambda_3$ is
consistent for small $g_2$.  The standard model gauge group is not
broken in this vacuum, which, for sufficiently small $g_2$, is
certainly the true minimum of the potential.

We have therefore found that our model's \ssb\ vacuum preserves the
standard model gauge group when $\Lambda_4>\Lambda_3\gg\Lambda_2$.
However, the regime of interest, $\Lambda_3>\Lambda_4\gg\Lambda_2$, is
not calculable. While we cannot prove that the above vacuum continues
to be stable (or sufficiently metastable) into this regime, it is
reasonable to assume that it does so for some range of
parameters, and that for $g_2\sim 1$ it leads to $F\sim \Lc^2$.
Nonetheless, since supersymmetry is restored for $y_A=0$, and the
dynamics of the theory tend to drive $y_A$ small at low energy, the
effects of this parameter deserve further discussion.

\subsection{Why $y_A$ is small}
\label{ssec:lightmess}

We claimed that $y_A$ is likely to be small, leading $\bar A, A$ to be
light.  In this section, taking an idealized limit, we clarify why
$y_A$ is driven small, and explain why we cannot estimate its size.
We also explain why we cannot determine a lower bound on $y_A$ from
the requirement that the global vacuum be phenomenologically
acceptable.

In particular, suppose that the couplings $g_3, \lambda_0,y_B,y_C$ are
all large at the Planck scale $m_P$ and chosen so that the theory
reaches the (conjectured) infrared fixed point discussed in section
\SSEC{suthree} at energies near $m_P$. Suppose also that
$g_4=g_i^{SM}=0$.  We do not know the anomalous dimensions
$\gamma_{\bar V},\gamma_A$, but the contribution from the couplings
$y_A,y_B$ are inevitably positive by unitarity (\SEC{betagammas}.) In
particular, $\gamma_{\bar V}$ need not be small in magnitude even when
$y_A$ is very small. Consequently $y_A$ is irrelevant (see
\Eref{ybeta}), and is driven small like a power (most likely less than
1) of the energy.
\be{yAmu}
y_A(\mu)\leq y_A(m_P) \left({\mu\over m_P}\right)^{\gamma_{\bar V}^*}
\ee
where $\gamma_{\bar V}^*$ is $\gamma_{\bar V}$ at the $SU(3)$ fixed
point (at which $y_A=0$.)

Accounting for $g_4,g_i^{SM}\neq0$ makes very little difference until
$g_4$ becomes substantial near $\Lambda_4$, making the anomalous
dimension of $\bar V$ smaller and probably negative.  The $Sp(4)$
physics will cause $y_A$ to change by some unknown factor $\kappa_4$,
probably greater than one.  As noted in section \SSEC{spfourbeta}, the
regime in which $g_4$ is large may be extended by a small beta
function, leading $\kappa_4$ to be larger than naively expected.  But
because the $SU(3)$ physics is nearly scale invariant, and $y_A$ is
negligibly small at low energies, it follows that for $\Lambda_4$ and
$\Lc$ much smaller than $m_P$, the confining physics of the $Sp(4)$
theory is independent of $\Lambda_4/m_P$.  In turn this implies that
$\kappa_4$ is independent of $\Lambda_4$ for small $\Lambda_4$.  The
low-energy value of $y_A$ is thus a monotonic decreasing function of
$\Lambda_4$:
\be{yAlow}
y_A(\Lc)\lsim y_A(m_P)
\left({\Lambda_4\over m_P}\right)^{\gamma_{\bar V}^*}\kappa_4 \ .
\ee

By varying the couplings at $m_P$, maintaining only
$\Lambda_3>\Lambda_4$, we may easily obtain any low-energy value of
$y_A$ that we want, as long as it is not very large.  We therefore
have no prediction for $y_A$ (especially as $\kappa_4$ cannot be
calculated,) and so we will always consider it a free parameter, which
we expect to be smaller than unity.

The one additional concern one might have is that since supersymmetry
is restored for $y_A=0$, due to the flat direction $A\bar V\bar V$
which breaks the standard model gauge group, perhaps the global
minimum of the potential will break the standard model gauge group if
$y_A$ is driven small.  However, as we will now see, it cannot be
shown that this occurs, and so no conclusions may be drawn.

Specifically, in the limit $y_A=0$, the model develops a classical
flat direction $A\bar V\bar V$.  This flat direction is not lifted
quantum mechanically, as can be seen through the following argument.
The expectation value for $\bar V$ gives mass to four triplets in $B$
and four antitriplets in $\bar T$, but leaves one triplet $B^5$
behind.  The $SU(3)$ gauge theory thus has three massless flavors
$q,B^5,\bar u,\bar d, C$.  This gauge group has a quantum modified
moduli space, and so at large expectation values the classical moduli
space is not modified.  The theory therefore has vacua at infinite
expectation values for $A,\bar V$ and the massless $SU(3)$ fields,
with a potential energy which is essentially flat at large vevs.

If $y_A$ is non-zero and small, however, a quartic potential for $A$
and $\bar V$ lifts these nearly flat directions.  With the assumption
of large $\vev{A}$ and $\vev{\bar V}$, there is no vacuum, even for
arbitrarily small $y_A$.  Therefore, $\vev{A}$ and $\vev{\bar V}$ must
be small; but will they be zero in the vacuum, as required
phenomenologically?  We can look for a minimum at very small
expectation values; in this case the confining of $Sp(4)$ and the
breaking of supersymmetry occurs as in \SSEC{dsbvac}, with the only
effect of small $y_A$ to make $A,\bar A$ massive.  Locally, then,
there is a minimum with $\vev{A}=\vev{\bar V}=0$.  Unfortunately, we
cannot be sure this is a global minimum; there could be a minimum for
$\Lc<\vev{\bar V}<\Lambda_3$, where strong coupling effects make a
reliable computation impossible.

Note that our model involves a slightly different situation.  We are
not assuming that $y_A$ is small at high scales, only that it may be
driven small at low energy.  The flat direction at large $\vev{\bar
V}$ is therefore strongly lifted.  Still, it is likely that there is a
lower bound on the low-energy value of $y_A$ below which the standard
model may be broken in the global minimum of the potential.  As this
bound cannot be calculated, and depends on $g_2$, we leave the
low-energy value of $y_A$ as a free parameter without a lower bound.

Is it possible, even when the standard model gauge group is conserved,
that the \ssb\ scale $F$ depends on $y_A$, and perhaps becomes much
less than $\Lc$ when $y_A\ll1$?  This seems unlikely to us.  We have
shown there is a acceptable \ssb\ vacuum whose value for $F$ is
controlled by $g_2$ [\Eref{MSdef}]; in the limit of small $g_2$,
finite $y_A$, this vacuum is the true minimum.  In the limit of small
$y_A$, finite $g_2$, the conjectured vacuum discussed in this section,
which breaks the standard model, might be the true minimum; its energy
density would be proportional to a power of $y_A$.  We would expect
there to be a first-order transition between these two vacua, and a
corresponding discontinuous transition in the functional dependence of
$F$ on $g_2$ and $y_A$.  In the allowed minimum of \SSEC{dsbvac},
the $y_A$ dependence of $F$ is sub-leading; conversely, if $F$
depends strongly on $y_A$, then the theory is probably in the
unacceptable vacuum.  Thus, although we cannot determine the range of
 $g_2$ and $y_A$ for which the theory prefers the acceptable vacuum, we
can assume that when it does so, the \ssb\ scale $\sqrt{F}$ depends very
little on $y_A$ and remains of order $g_2^{3/14}\Lc$.

\subsection{The $SU(5)$ global symmetry and coupling constant unification}
\label{ssec:sufive}

We have added extra matter to the standard model, and thus run the
risk that we will drive the standard model couplings to a Landau pole
below the string or Planck scale.  Furthermore, we have added
additional interactions which do not exactly satisfy $SU(5)$ relations
and which can become strong.  In this section we confirm that
unification at finite coupling is possible, and that $SU(5)$ can be
preserved as an approximate global symmetry, provided that there
are no strong violations of $SU(5)$ at the Planck scale.

First, we consider the limit in which $SU(5)$ is preserved by the
superpotential \Eref{fullsup}, and we verify that the standard model
couplings can unify at finite coupling.  We have added five triplets
and antitriplets to $SU(3)^{SM}$ in Table 1, which change the
perturbative QCD beta function coefficient $b_0$ from $+3$ to $-2$.
These fields disappear from the theory at the scale $\Lc\sim 100
\TeV$, except for the fields $\psi_A,\psi_{\bar A}, A$.  A one-loop
analysis indicates this is a borderline case, while a two-loop
analysis ignoring the Yukawa couplings in \Eref{fullsup} would suggest
that the standard model gauge couplings hit a Landau pole below the
Planck scale.  However, the situation is modified by the Yukawa
couplings.

 In particular, the QCD beta function above $\Lc$ reads
\be{QCDbeta}
\beta_{g_3^{SM}} = -{(g_3^{SM})^3\over 16 \pi^2}
{-2 + {3\over 2}\gamma_A  + {3\over 2}\gamma_B
+ 2\gamma_{\bar V}\over
1-{3(g_3^{SM})^2\over 8\pi^2}} \ .
\ee
(For simplicity of discussion, we ignore in this expression the
anomalous dimensions of standard model fields, as generated by
standard model gauge and Yukawa couplings.)  At energy scales well
above $\Lc$ but below $\Lambda_3$, where $g_3$ is large, the
expectation is that $\gamma_A$ is positive, $\gamma_{\bar V}$ is
positive, and $\gamma_B$ is negative. If $y_B$ were small,
$\gamma_{\bar V}$ and $\gamma_A$ would be very small and $g_3^{SM}$
would run faster than the one-loop analysis would suggest.  However,
because $y_B$ is large, and because $y_A$ may be large at high scales,
$\gamma_A$ and $\gamma_{\bar V}$ need not be small, and likely make
$g_s^{SM}$ run more {\it slowly} than the one-loop analysis.  Once we
approach $\Lc$ and $g_4$ becomes large, the analysis is even less
under control; no argument can be constructed indicating either that
$g_3^{SM}$ must run faster or slower than perturbatively indicated.
Finally, below $\Lc$, the QCD beta function becomes negative, although
the scalar $A$ and the fermions $\psi_A, \psi_{\bar A}$ will reduce
the QCD beta function slightly below its MSSM value.  Altogether the
uncertainties in the anomalous dimensions prevent us from
demonstrating unambiguously that $g_3^{SM}$ remains finite, but since
the theory at one loop is marginally acceptable, there likely exists a
region of parameter space in which this is the case.

Suppose that $g_3^{SM}$ does not reach a Landau pole; what about the
other standard model couplings?  It is well-known that addition of
complete $SU(5)$ multiplets to the standard model does not ruin
coupling constant unification.  This is true at one loop in the
standard model couplings and to all orders in other couplings.  The
proof in a supersymmetric theory is direct.  As seen in \Eref{SVbeta},
to leading order in a weak coupling constant the beta function is
proportional simply to $g^3$ times $3C_2(G_k)-\sum_p
T(\phi_p)[1-\gamma_{\phi_p}]$.  The usual statement of coupling
constant unification is that a complete $SU(5)$ multiplet $\{\phi_j\}$
preserves unification because $\sum_j T(\phi_j)$ is the same for each
standard model group factor, leading to equal shifts in $b_0 =
3C_2(G_k)-\sum_p T(\phi_p)$ for the three groups and preserving both
unification and the unification scale.  In this case, the $SU(5)$
multiplets have large anomalous dimensions due to strong interactions
involving the 4-3-2 sector. However, since the fields $\{\phi_j\}$ all
have the same anomalous dimension $\gamma_{\{\phi_j\}}$ (by
approximate $SU(5)$ flavor symmetry) the sum $\sum_j
T(\phi_j)[1-\gamma_{\phi_j}]= [1-\gamma_{\{\phi_j\}}]\sum_j T(\phi_j)$
is essentially the same in each standard model group factor.  Again,
unification is perserved.

Thus, if $g_3^{SM}$ does not hit a Landau pole, neither will
$g_2^{SM}$ or $g_1^{SM}$.  Furthermore, coupling constant unification
will be preserved despite the strong coupling effects.

Now, we consider the possiblity that $SU(5)$ is broken in the
superpotential \Eref{fullsup}.  In this case the multiplets $\bar V,
A, B$ must be broken up into their component multiplets under the
standard model gauge group, each with its own anomalous dimension. As
an simplified example, suppose that $y_A$ is very small and that we
can ignore it and the field $A$.  Consider the fields $B,\bar V$,
which contains as submultiplets a color triplet and antitriplet
$B_3,\bar V_3$ and weak isodoublets $B_2, \bar V_2$.  Let us consider
the effect on the coupling $y_B B\bar T\bar V$, which now becomes two
couplings $y_{3} B_3\bar T\bar V_3 + y_{2} B_2\bar T\bar V_2$.
The anomalous dimensions of the relevant fields are
given at one loop by
%
\be{BVgammas}\begin{array}{rclrcl}
16 \pi^2 \gamma_{B_2}&\approx& 2
y_{2}^2 - {16\over 3} g_3^2  &
16 \pi^2 \gamma_{B_3}&\approx& 2
y_{3}^2 - {16\over 3} g_3^2 \\ \\
16 \pi^2 \gamma_{\bar V_2}&\approx& {3\over 2}
y_{2}^2 - {15\over 2} g_4^2 &
16 \pi^2 \gamma_{\bar V_3}&\approx& {3\over 2}
y_{3}^2 - {15\over 2} g_4^2 \\ \\
16 \pi^2 \gamma_{T} &\approx& {1\over 2}
(3y_{3}^2 +2y_{2}^2) - {16\over3} g_3^2 - {15\over 2} g_4^2 \ .
\end{array}
\ee
The beta functions for $y_{2}$ and $y_{3}$ are
\be{ybbetas}
\beta_{y_{2}} =
\half y_{2}\left(\gamma_{B_2} + \gamma_{\bar V_2} + \gamma_{\bar T}\right)
 \ ; \
\beta_{y_{3}} =
\half y_{3}\left(\gamma_{B_3} + \gamma_{\bar V_3} + \gamma_{\bar T}\right)
\ee
Consider the ratio $r=y_{2}/y_{3}$; its beta function
can be written
\be{rbeta}
{1\over r}{\partial r\over \partial \mu}
= (\gamma_{B_2} -\gamma_{B_3}+
 \gamma_{\bar V_2}- \gamma_{\bar V_3} )
\approx {7\over 64\pi^2} y_{2}y_{3}\left(r-{1\over r}\right)
\ee
Thus, if the product of the Yukawa couplings $y_{2}$, $y_{3}$ is
small, both couplings will grow with the ratio $r$ remaining fixed.
However, the effect of the Yukawa couplings on the $r$ beta function
will cause $r$ eventually to relax toward one.  We see then that when
the $y_B$ couplings become large, as we expect them to be at low
energy, $SU(5)$ violation tends to be driven small.

 A similar analysis shows that the couplings denoted $y_A$ are also
driven toward $SU(5)$ universality if they are large, though not if
they are small.  Either way, the effects of $SU(5)$ violation are not
large and will not prevent unification of standard model gauge
couplings.

In conclusion, this model probably allows the unification of standard
model couplings.  All strong couplings will be nearly
$SU(5)$-preserving as a result of strong dynamics; weak couplings may
violate this global symmetry.  We will see some physical consequences
of this symmetry below.

\section{Below the Scale of Supersymmetry Breaking}
\label{sec:phenomena}

In this section we discuss various predictions and interesting
features of the model which are relevant for energies in the TeV
region and below, including mass relations between sfermions and the
$A,\psi_A,\psi_{\bar A}$ fields, large superoblique corrections, and
possible decay modes for the $A,\psi_A,\psi_{\bar A}$ particles.

\subsection{Spectrum of the light non-standard-model fields}
\label{ssec:AAbar}

It is helpful first to consider the limit where standard model gauge
couplings and $y_A$ are taken to be arbitrarily small (ignoring the
appearance of a \susy-preserving vacuum at $y_A=0$.)  All the standard
model fields and $A,\psi_A$ are decoupled.  The only interacting light
fields are $\psi_{\bar A}$ and the light fields in the \ssb\ 3-2
sector.  Here we discuss their interactions and properties.

For simplicity, we will assume that the 3-2 sector, which breaks
\susy, does not break its U(1)-``hypercharge'' flavor symmetry.  This
is true for small $\lambda_0$ \cite{threetwo} but may not be true for
$\lambda_0$ large.  With unbroken hypercharge, the 3-2 sector has two
light particles after supersymmetry breaking: the goldstino $G$, which
is eaten by the gravitino and obtains a mass $F/(\sqrt{3}m_P)$, and a
fermion $\eta$ required to saturate the ``hypercharge'' anomalies. We
will refer to the superfield containing $\eta$ as $S_\eta$, and the
superfield containing the Goldstino as $X$.\footnote{Note that the
F-term of $X$ is just $F$, the Goldstino decay constant.}  If instead
the hypercharge symmetry is broken, there will be a massless Goldstone
boson to replace $\eta$. However, the difference is irrelevant for
present purposes, as its effect on phenomenology of
standard-model-charged particles is limited to the decays of the heavy
fields $B,\bar B$, which have mass of order $\Lc$.

Before getting into the discussion of the specific scalar masses in
our model, we note that \ssb\ scalar mass-squared terms are of two
types, ``holomorphic'' and ``nonholomorphic'', that are of differing
size in the limit of small supersymmetry breaking. Holomorphic scalar
mass terms are defined to be those which couple scalar fields of the
same chirality. In the limit $F\ll\Lc^2$, supersymmetry-breaking
nonholomorphic scalar mass terms for a generic strongly coupled
superfield $\Phi$ come from terms in the K\"ahler potential of the
form
\be{nonholomass}{16\pi^2\over\Lc^{2}}\int d^4\theta\
\Phi^{\dagger}\Phi X^{\dagger} X\  ,
\ee
which will give $\Phi$ a nonholomorphic (of type $\Phi\Phi^\dagger$)
\ssb\ mass-squared of order $(4\pi F/\Lc)^2$.  Holomorphic
supersymmetry-breaking scalar mass terms (of type $\Phi^2$) can also
be induced. To see this, one can minimize the scalar potential and
solve for the $\Phi$ auxiliary field in the presence of K\"ahler terms
in the effective theory such as
\be{holomass}
{4 \pi\over \Lc}\int d^4\theta\
\Phi^{\dagger}\Phi (X^{\dagger}+ X)\  ,
\ee
and effective superpotential terms such as
\be{holomasssuper}
\int d^2\theta\
m \Phi^2\  + {\rm h.c.}.
\ee
One then finds a scalar mass-squared term in the potential
\be{hmterm}\sim{4\pi m\over\Lc} F \Phi^2 + {\rm h.c.}\ee

The superfield $S_\eta$ is a participant in the strong coupling
dynamics of the 3-2 sector. Naively one might guess that its scalar
component gets a \ssb\ mass-squared of order $F$.  In fact, a more careful
analysis, using a supersymmetric effective Lagrangian, shows that for
$\F\ll\Lc$ its mass is much smaller than this, because the unbroken
U(1) symmetry prevents the scalar from getting a holomorphic mass
term.  The mass-squared of the $S_\eta$ scalar thus gets only a
nonholomorphic contribution, of order $({4\pi F}/\Lc)^2$.

The field $\bar A$ is composite at the scale $\Lc$, and so its scalar
component also gets a nonholomorphic mass-squared $m_{\bar A}^2$ of
order $(4\pi F/\Lc)^2$.  We will {\it assume} this mass-squared is
positive; since it is induced through strong coupling it cannot be
computed.

Now let us consider turning on $y_A$.  This gives the $A,\bar A$
multiplets a supersymmetry preserving mass, of size $ m_\Psi\sim
|y_A|\Lc/(4\pi)$.  In particular, $\psi_A$ and $\psi_{\bar A}$ become a
Dirac fermion $\Psi_A$.  Also induced are a holomorphic \ssb\
mass-squared $m_{AH}^2$ and a nonholomorphic \ssb\ mass-squared
$m_A^2$ for the scalar $A$.  Their sizes are set by the following
consideration: all supersymmetry-breaking interactions involving $A$
are mediated through its coupling to $\bar A$, and are therefore
suppressed by one power of $y_A/(4\pi)$ for each $A$ on an external
leg.  We find therefore that $m_{AH}^2 \sim y_A F$ and $m_A^2\sim (y_A
F/\Lc)^2\sim (y_A/4\pi)^2 m_{\bar A}^2$.

Next, when the standard model gauge couplings are made non-zero, the
gauge bosons lead to a conventional positive gauge-mediated
contribution $m_{GM}^2$ to the $A,\bar A$ scalar masses-squared, of
order $(\alpha^{SM}_k F/\Lc)^2$.  The $A, \bar A$ scalar mass-squared
matrix thus has the form
\be{amassmat}
\bordermatrix{&A&{\bar A}^*\cr
& & \cr
A^*&{m_\Psi}^2+ m_A^2+m_{GM}^2
& m_{AH}^2\cr &&\cr
\bar A & m_{AH}^2&  {m_\Psi}^2+m_{\bar A}^2+m_{GM}^2\cr}\ .
\ee
For small $y_A$, one linear combination of the scalars, which is
mostly $A$, is relatively light.

The experimental signatures of the $\Psi_A$ and $A$ scalar depend on
which one is lighter. When $y_A$ is sufficiently small, \ie\ for
$y_A\Lc/(4\pi)\ll \alpha^{SM}_k F/\Lc$, the fermion $\Psi_A$ is
lightest member of the $A,\bar A$ multiplet. Otherwise, it is not
possible to say which is lighter.

Note that the masses of the $B,\bar B$ multiplet have a similar form
to those of $A,\bar A$, except that we expect $y_B$ to be large, and
so none of these particles will be light.

\subsection{The Message of Supersymmetry Breaking}
\label{ssec:BBar}

In conventional models of GMSB, violations of supersymmetry in the
MSSM sector can be reliably computed from diagrams containing loops of
particles carrying ordinary gauge charges. In our model, one might be
tempted to compute superpartner masses by considering loops containing
the $A,\bar A$ and $B,\bar B$ fields. However, unless a theory has
messengers with a canonical K\"ahler potential and a mass-squared
matrix with vanishing supertrace, the squark and slepton masses come
from ultraviolet divergent diagrams and are sensitive to the
high-energy, strongly-coupled physics. The squark, slepton and gaugino
masses are computable in the limit $\Lambda_4\gg\Lambda_3 >\Lambda_2$
and $y_A, y_B, y_C, \lambda_0\ll 1 $.  In this limit, all the fields
carrying standard model gauge charges are effectively weakly coupled
below the scale $\Lambda_4$.  By contrast, our model has some large
Yukawa couplings and $\Lambda_3>\Lambda_4$.  Here, the situation is
not so straightforward, even if $y_A$ and $y_B$ are small and the
$A,\bar A, B,\bar B$ messengers are much lighter than $\Lc$.

To see the limitations of perturbative computation, consider the limit
$y_A, y_B\ll1$ and $\F\ll |y_A|\Lc, |y_B|\Lc\ll\Lc$. In this case the
nonholomorphic mass terms for the fields $A, \bar A, B,\bar B$ are
suppressed, as is the supertrace of their mass matrices, and all
members of these supermultiplets appear as effectively weakly-coupled
fields below the scale $\Lc$, with masses of order $y_A \Lc/(4\pi),
y_B \Lc/(4\pi)$.  As we discussed in the previous section, holomorphic
supersymmetry-breaking scalar mass-squared terms $m_{AH}^2$, $m_{BH}^2$, of
type $A\bar A$, $B\bar B$ respectively, appear in the low-energy
effective theory, with $m_{AH}^2\sim y_A F$ and $m_{BH}^2\sim y_B F$.
The nonholomorphic supersymmetry-breaking scalar mass terms are of
order $(4\pi F/\Lc)^2$ and are relatively suppressed.  In this limit,
which we will call the ``holomorphic'' limit, the contributions of the
$A$ and $B$ messengers to the ordinary superpartner masses are
positive and of order $\alpha_k^{SM} F/\Lc$, and can be perturbatively
computed once their masses are known. Note that these contributions
are independent of $y_A$ and $y_B$.  But this is not the whole story.
Near the scale $\Lc$, there could also be a host of broad resonances,
with standard model quantum numbers and supersymmetry-breaking
holomorphic mass-squared terms of order $4\pi F$, which give an
equally large contribution to the superpartner masses. Similar
conclusions can be reached by applying naive dimensional analysis to
graphs involving all the fundamental strongly-coupled fields.  It is
therefore not appropriate to regard the $A,\bar A$ and $B,\bar B$
superfields as the only messengers --- the message is carried by the
supersymmetry breaking sector as a whole.

We do not expect the holomorphic limit to apply, however.  We expect
that $y_A$ is small, so the holomorphic mass-squared term $m_{AH}^2$,
which is proportional to $y_A$, does not dominate over the
nonholomorphic term $m_{\bar A}^2$, which is proportional to $(4\pi
F/\Lc)^2$.  When the supertrace of the messenger masses-squared does
not vanish, the contribution of the messengers to the squark and
slepton masses-squared is logarithmically divergent in the low-energy
effective theory. This divergence is cut off in the full theory, but
logarithms of the ratios of messenger masses to $\Lc$ can appear in
the squark and slepton masses.

By considering the contribution of loops containing the $A,\bar A$
multiplet to ordinary sfermion masses-squared
\cite{PoppitzTrivedi,nmsb}, we can show that the regime with
$(y_A/4\pi) \Lc^2\ll 4 \pi F\ll \Lc^2$ is ruled out because of a large
negative logarithm. We will refer to this region of parameter space
as ``log-dominated''.  A perturbative calculation of the loop
contribution from the $A, \bar A$ multiplet to squark and slepton
masses-squared\cite{PoppitzTrivedi,nmsb}, for $(y_A/4\pi) \Lc^2\ll 4
\pi F$, is of order ${\alpha^{SM}_k}^2(F^2\Lc^2)\log(4\pi F/\Lc^2)$
and is negative. The uncertainty in this calculation is of the same
size as the effects from the rest of the strongly interacting sector and
is of order ${\alpha^{SM}_k}^2(F^2/\Lc^2)$.  For $4 \pi F\ll \Lc^2$
the log enhances the negative contribution of the $A, \bar A$
multiplet to squark and slepton masses-squared. Since there is no
other logarithmically enhanced contribution, we conclude that the
log-dominated regime is excluded.

For $4\pi F\sim\Lc^2$, we are not in the log-dominated limit, and
there are equally large contributions to squark and slepton
masses-squared of unknown sign. We refer to the parameter region with
$4\pi F\sim\Lc^2$, $y_A\ll 4\pi$, a natural one for our model, as the
``light messenger'' regime.  We will assume that the model has an
acceptable region of parameter space in the light messenger regime
with positive squark and slepton masses-squared.  Note that all
contributions which are not suppressed by weak couplings are
approximately $SU(5)$ symmetric, so the two-loop contribution to all
standard model sfermion masses-squared has the same sign.

In such an acceptable regime, estimates for gauge-mediated \ssb\
masses can be made following the usual arguments.  For a sfermion $\tilde
f_r$ in a representation $r$ under the standard model, we find
\be{smasses}
\tilde m^2_{\tilde f_r} = c_m \left(\sum_{k=1}^3 C_{rk}
{\alpha_k^{SM}}^2\right)
{F^2\over\Lc^2} \ .
\ee
Here the $C_{rk}$ are the Casimir indices for the representation $r$
of the standard model gauge group $[S]U(k)$, while $c_m$ is an unknown
constant, assumed positive.  Because sfermion masses-squared scale
with the effective number of messengers, and perturbatively we have
the equivalent of five messenger $5+\bar 5$'s of global SU(5), we
estimate $c_m$ is of order 5, and, due to approximate $SU(5)$
symmetry, is approximately independent of the representation $r$.  For
the standard model gauginos $M_i$, we expect
\be{gauginomasses}
M_i =
d_m\alpha_k^{SM}{F\over\Lc}
\ .
\ee
Here $d_m$ is another unknown constant, also of order 5.

The fields $A$ and $\bar A$ have quantum numbers of a ${\bf 10}$ and
${\bf \overline{10}}$ under $SU(5)$.  However because the global
$SU(5)$ is broken by weak couplings such as the $y_A$, the superfields
$A,\bar A$ are each three multiplets with different masses, $A_r$,
$\bar A_{\bar r}$ where $r=(3,2)_{1/6},(\bar 3,1)_{-2/3},(1,1)_{1}$
labels the representation of $A_r$ under the standard model.  We
assume that the three different $y_A$ couplings are all of the same
order.

If the $y_{A_r}$ are of order one, none of these particles will be
observable in the near term.  But if the $y_{A_r}$ are small, then the
fields $\Psi_A,A$ might be light enough to be found soon. The
color-neutral fields are likely to be the lightest since the scalar
masses do not receive a large gauge-mediated contribution from color,
and because the renormalization group predicts enhancement of the
Yukawa couplings for colored particles. Since $\Psi_A$ and $A$ make up
3/2 of an $SU(5)$ supermultiplet, non-observation of $\bar A$ in the
near vicinity would suggest that $\bar A$, and the $\psi_{\bar A}$
component of $\Psi_A$, are participants in the
supersymmetry-breaking dynamics.  One might then probe this dynamics
by studying the interactions of the $\Psi_A$ particle.

The $SU(5)$ global symmetry could be broken by the superpotential
couplings. We have assumed (see \SSEC{sufive}) that all such couplings
either are weak or, if strong, are drawn towards an $SU(5)$ invariant
fixed point. This assumption may be tested via the relations of
eqs. \eref{smasses} and \eref{gauginomasses}. Another interesting test
is possible if the $\Psi_A$ and $A$ masses are measured. Global
$SU(5)$ symmetry gives the sum rule
\be{sumrule}
m_{A_r}^2- \tilde m^2_{\tilde f_r} = x m_{\Psi_r}^2\ ,
\ee
where $x$ is a constant independent of $r$.

In addition, our model (in contrast to most GMSB models) may generate,
through strong dynamics, relatively large ``superoblique'' corrections
\cite{superobliquea,superobliqueb} --- supersymmetry-violating
contributions to the gaugino couplings. The typical expected size of
such corrections can be estimated from naive dimensional analysis to be
\be{superoblique}
{g_i^{SM}-{\tilde{g}}^{SM}_i\over g^{SM}_i}\sim {\alpha^{SM}_i\over4\pi}\ ,
\ee
where the ${\tilde{g}}_i$ are the gaugino Yukawa couplings.  In the
light messenger limit, a logarithmic
enhancement of this contribution can be reliably computed to be
\cite{superobliquea,superobliqueb}
\be{superobliquelog}
{g_i^{SM}-{\tilde{g}}^{SM}_i\over g^{SM}_i}={\alpha^{SM}_i\xi_i\over 4\pi}
\log\left({4\pi F\over\Lc m_{\Psi}}\right)\ ,
\ee
where $\xi_i= (5/3),1,1$ for $i=1,2,3$ respectively, and we have
neglected differences between the different $\Psi_r$ masses.

\subsection{Trilinear scalar terms and the Higgs sector}
\label{ssec:muterm}

As in most gauge-mediated supersymmetry models, the supersymmetry
breaking parameter $B$ which governs mixing of the $H_u$ and $H_d$
scalars, and the trilinear terms among scalar fields of the MSSM, are
generated at two loops, and are relatively small, of order
$(\alpha^{SM})^2 F/(4\pi \Lc)$.  The only way to generate such terms
at one loop is to have heavy gauginos which couple to MSSM particles,
such as when the standard model gauge group is embedded in a larger
group.

Our model does not address the generation of the $\mu$ term, i.e. the
supersymmetric Higgsino mass parameter in the MSSM superpotential.  A
long standing problem for supersymmetric theories is to explain why
this parameter should be of order the weak scale, as is
phenomenologically required \cite{GFM}.  Several solutions to this
problem, which were proposed in the original models of GMSB with DSB
\cite{newGMSBa,newGMSBb,newGMSBc}, would also work for the present
model.  Basically these solutions rely on generating the $\mu$
parameter from the vev of a fundamental singlet which is coupled to
the MSSM sector.  Since these original models, several newer solutions
have been proposed. However none of these newer solutions
\cite{GFM,munewa,munewb,munewc} will work for the present model,
either because they need $\F$ to be substantially larger than $\sim
10^5$ GeV, or because they require fundamental gauge singlet
superfields with renormalizable couplings to the messenger sector.
Witten's sliding singlet mechanism \cite{sliding} has been claimed to
generate an acceptable $\mu$ term in some models with $\F\sim
10^5$~GeV \cite{CF}; however, for this mechanism to generate an
acceptable $\mu$ parameter, the $B$ parameter must also be large,
which is not straightforward to arrange in this model.

\subsection{Decay of the messengers}
\label{ssec:messdecay}

So far we have assumed that there are no superpotential couplings
between the MSSM and the supersymmetry-breaking sectors.  This
assumption could lead to stable charged particles in the messenger
sector. Such particles are easily ruled out via, \eg\, searches for
heavy hydrogen.

The heavy $B\bar B$ messengers can decay via an $Sp(4)$ instanton into
two $A$ messenger particles and a neutral $C$ messenger. The $C$
messengers are allowed to decay via the couplings $h_1,h_2$ into the
light particles of the 3-2 sector, i.e. the Goldstino and the massless
fermion mentioned in section \SSEC{AAbar}. However, the lightest $A$
messengers in each charge sector are stable unless new couplings are
introduced.

The easiest way to eliminate the cosmological problems of stable
charged messengers is to assume that dimension-five couplings between
the supersymmetry breaking and MSSM sectors are allowed, \eg\ $ A \bar
U \bar D \bar E$, (where $\bar U, \bar E$ are respectively the MSSM up
antiquarks and charged antileptons). Even if suppressed by the reduced
Planck scale, dimension 5 couplings lead to lifetimes for messengers
with TeV scale masses (such as most of the $A$ multiplet might have)
of around $10^{-2}$ seconds. In this case the lightest messengers
would be irrelevent for cosmology but might be detectable in a
collider experiment as heavy, long-lived charged particles.

Alternatively, (if $A$ is in the ${\bf 10}$, not the ${\bf \overline{10}}$
representation) one could allow the following renormalizable couplings
which are consistent with baryon and lepton number conservation and
all the gauge symmetries\footnote{Note that the simple alternative of
allowing the $A$ particles to mix with those ordinary quarks and
leptons with the same quantum numbers leads to rapid proton decay. We
can impose a discrete symmetry to forbid such couplings and still
allow those of eq.~\eref{decaysupII}.}.  Such couplings, if larger
than $\sim 10^{-6}$ or so, will allow prompt decays of all new heavy
particles.
\be{decaysupII}
\lambda^{ij}_{LL}L_iL_j A_{(1,1)_{1}} +\lambda^{ij}_{LD}L_i\bar
D_jA_{(3,2)_{1/6}}
+\lambda^{ij}_{DD}\bar D_i\bar D_j A_{(\bar 3,1)_{-2/3}} \ .
\ee
Here $i,j=1,2,3$, and $\bar D_i, L_i $ are the left chiral superfields
for respectively the down antiquarks and the lepton doublets, and the
coupling to the $A$ field is to the appropriate component such that
the coupling is gauge invariant. Hence when
$\lambda_{DD},\lambda_{DL},\lambda_{LL}$ are nonzero, $A$ contains
superfields with the quantum numbers of a diquark, a leptoquark, and a
dilepton.

The addition of the $\lambda_{LD}, \lambda_{LL}, \lambda_{DD}$ terms
allows for a new classically flat direction parametrized by the
superfields $\bar V^2, \bar D, \bar L $, along which the $Sp(4)$ gauge
symmetry is completely Higgsed. Along this flat direction the $Sp(4)$
dynamics no longer lifts the classically flat directions parametrized
by $q^2 B$, $B^3$, and the $SU(3)\times SU(2)$ factor is completely
Higgsed as well. Thus there is a classically flat direction along
which nonperturbative gauge effects are small, and so the couplings of
\eref{decaysupII} lead to a new, supersymmetric vacuum at infinity in
field space. Still, provided the couplings are sufficiently small,
there will still be a local supersymmetry-breaking minimum in the
vicinity of the minimum which is there in the absence of such a
coupling. Furthermore, when $\lambda_{LD}, \lambda_{LL}, \lambda_{DD}$
are small, the barrier between the desired and the new minimum becomes
very large and the lifetime of the desired vacuum becomes much longer
than the lifetime of the observed universe. There is no known reason
why we should not be living in such a false vacuum.

Another reason for requiring any couplings between $A$ and the MSSM
fields to be small is that such couplings can lead to flavor-changing
neutral currents at tree level. The strongest constraint is on certain
combinations of the couplings
$\lambda_{DL}^{11},\lambda_{DL}^{22},\lambda_{DL}^{12}$ and
$\lambda_{DL}^{21}$, since the scalar leptoquark exchange can give a
tree level contribution to rare $K$ decays such as $K_L\to \mu e$ and
$K\to \mu e \pi$. Furthermore, as argued in \SSEC{lightmess}, the
colored $A$ scalars may be as light or lighter than the squarks. The
leptoquark contribution to rare $K$ decays will be compatible with all
current bounds provided the $\lambda_{DL}$ couplings are all smaller
than $\sim 10^{-3}$.
Fortunately it is natural for these couplings to
be small since they are irrelevant in the energy regime between
$\Lc$ and $\Lambda_3$.

We conclude that another acceptable scenario is that the couplings
$\lambda^{LD}, \lambda^{LL}, \lambda^{DD}$ are all present but very
small.  In this case the only non-standard stable or long-lived
particles in the model are the gravitino $\G$ and the massless fermion
discussed in \SSEC{AAbar}, and neither of these lead to any
phenomenological problems or cosmological difficulties.  In this
scenario, the lightest messengers (the Dirac fermion $\Psi_A$ and the
complex scalar $A$) might be pair produced and their decays observed
in a hadron collider. The scalars are R parity even.  The scalar
dilepton, which decays into a charged lepton and a neutrino, could be
the lightest of the even R parity messengers, potentially as light as
the right handed charged sleptons.  The leptoquark scalars are a
nearly degenerate weak doublet, with the charge $2/3$ member decaying
promptly into charged lepton and a down quark jet, and the charge
$-1/3$ member into a down quark jet and a neutrino.  These could be as
light or lighter than the ordinary squarks.  Such leptoquarks have
already been searched for at Fermilab
\cite{leptoquarksearcha,leptoquarksearchb} but will have so far
escaped detection if heavier than 225 GeV.  The $\Psi_A$ fermions
could be lighter than their scalar superpartners. If heavy enough, the
messenger fermions decay into an ordinary quark and lepton and a
squark or a slepton. If such decays are not kinematically allowed they
must decay via virtual squarks and sleptons. In either case, the decay
chain for a messenger fermion always leads to the NLSP and its typical
decay signature.

\section{Conclusions}

We have presented an explicit example of gauge-mediated dynamical
supersymmetry breaking with a single supersymmetry breaking and
messenger scale in the 10--30 TeV region.  Our model has no fine
tuning, explicit mass parameters, or {\it ad hoc} small numbers.  The
model is relatively simple, but exhibits a rich variety of phenomena.
We have discussed the dynamics and properties of this model in detail.
This requires a careful analysis of the effects of strong coupling
that goes well beyond the use of holomorphy to constrain the
superpotential.

While this model preserves the usual successes (no flavor changing
neutral currents, a predictable spectrum) and difficulties (the $\mu$
and $B\mu$ problems) of gauge-mediated supersymmetry breaking, it has
some unusual features and predictions as well.  Some of these are
central to the model and would be typical of any theory of this type.
\begin{enumerate}
\item All couplings in the theory are of naturally of order one at
 high scales, although some are forced to be very large or very small
 at low-energy by the effects of strong dynamics.
\item Despite the large contribution of new strong interactions to the
  beta functions of the standard model gauge couplings, the usual
  supersymmetric GUT relations are protected by an approximate SU(5)
  global symmetry.
\item Although the theory has two strongly coupled gauge groups, a
 single dynamical scale controls the physics of the \ssb\ and
 messenger sectors.  This is due to a natural approximate fixed point,
 whose dynamics washes out the effects of the other energy scales.
\item As a direct result of the lack of segregation between
  supersymmetry breaking and messenger dynamics, the \ssb\ scale is
  low, resulting in an NLSP which can decay promptly into an
  ordinary particle and a gravitino.
\item The messenger sector is completely chiral with respect to the
  underlying gauge symmetries; it only becomes vectorlike after one of
  the gauge groups undergoes a confining transition.  The confining
  transition sets both the \ssb\ scale and the mass scale of the
  messenger sector.
\end{enumerate}

Somewhat more model-dependent but still reasonably generic properties
depend on the irrelevance, as a result of strong dynamics, of a
certain coupling in the superpotential. This coupling may be driven
much less than one, and if it is small (between $10^{-3}$ and
$10^{-1}$) it leads to a number of interesting effects.
\begin{enumerate}
\item Certain messenger chiral superfields only couple weakly to the
 \ssb\ dynamics, while most of them are strongly coupled.  This leads
 some of the particles in the messenger/DSB sector to have
 \susy-preserving masses much smaller than the \ssb\ scale.  These
 ``light messengers'' might be discovered soon by hadron colliders.
\item The chiral structure of the model leads the light messenger
 supermultiplets to have \ssb\ mass-splittings which differ
 substantially from those of their complex conjugates. Consequently,
 the light messenger particles do not form a set of complete
 supermultiplets.
\item The large mass splittings in the light messenger supermultiplets
 cause ``superoblique'' radiative corrections to the gaugino couplings
 to be logarithmically enhanced, and thus much larger than in
 weakly-coupled gauge mediated models.
\item The usual constraints on light messengers, due to their tendency
 to give negative mass-squared to standard model fermions, are evaded
 as a result of the strong-coupling dynamics in the \ssb\ sector.
\item The $SU(5)$ global symmetry gives a sum rule for the light
  messenger masses.
\end{enumerate}
If indeed there are light messengers (which is the most natural regime
for the theory,) these effects make the model experimentally
distinguishable from both gravity-mediated models and other proposed
gauge-mediated models.

We hope that the novel features of this model will be
thought-provoking, and will stimulate further research into the role
that strong gauge dynamics may play in the process of supersymmetry
breaking.

\bigskip
We would like to thank Yuri Shirman for useful
discussions. A. N. would like to acknowledge the Aspen Center for
Physics for hospitality during the inception of this work.

   \nocite{*}                
   \bibliography{432}        
\bibliographystyle{h-physrev}
\end{document}